\documentstyle[aps,pre,multicol,epsf]{revtex}
\def\be{\begin{equation}}
\def\ee{\end{equation}}

\def\bea{\begin{eqnarray}}
\def\eea{\end{eqnarray}}

\begin{document}

\title{Liesegang patterns: Effect of dissociation of the invading electrolyte}
\vspace {1truecm}

\author {B. Chopard${}^1$, M. Droz${}^2$, J. Magnin${}^2$, 
Z. R\'acz${}^{3}$, and M. Zrinyi${}^4$}
\address{{}$^1$ D\'epartement d'Informatique, University of Geneva, CH-1211 
Gen\`eve 4, Switzerland}
\address{{}$^2$ {D\'epartement de Physique Th\'eorique, Universit\'e de 
Gen\`eve,
CH 1211 Gen\`eve 4, Switzerland.}}
\address{{}$^3$Institute for Theoretical Physics,
E\"otv\"os University,
1088 Budapest, Puskin u. 5-7, Hungary}
\address{${}^4$Department of Physical Chemistry, Technical University of 
Budapest, H-1521 Budapest, Hungary}
\date{September 29, 1998}
\maketitle
\begin{abstract}

The effect of dissociation of the invading electrolyte on the 
formation of Liesegang bands is investigated. 
We find, using organic 
compounds with known dissociation constants, that 
the spacing coefficient, $1+p$, that characterizes 
the position of the $n$-th band as $x_n\sim (1+p)^n$, decreases 
with increasing dissociation constant, $K_d$.
Theoretical arguments are developed to explain   
these experimental findings and to calculate explicitly the 
$K_d$ dependence of $1+p$.
\end{abstract}

\maketitle
\vspace{0.5truecm}
\begin{multicols}{2}

\section{Introduction}

Liesegang patterns are  quasi-periodic  structures  precipitated 
in the wake of a moving reaction front. Although they appear only
in specific  physico-chemical conditions, these structures  
are widespread in nature and can be found in systems ranging 
from biological  
(populations of bacteria) to geological 
(structures in agate rocks)~\cite{{liese},{Henisch}}.
Depending on the geometry and the dimensionality of the system, 
the observed patterns are bands, rings 
or spheres, although it is also possible to generate more particular 
patterns such as spirals.

In chemistry, such structures are produced by allowing two chemicals, 
called  $A$ and $B$, initially separated, to mix through diffusion. 
The two components are chosen so that they react and form  
a nonsoluble product, called precipitate. 
In a typical example, $B$ (the inner electrolyte) is initially 
dissolved in a gel and placed in a test tube. 
At time $t=0$, 
the reagent $A$ (outer electrolyte) is poured into the tube and
it starts to diffuse into the gel. 
As the result of chemical and physical mechanisms 
such as nucleation, aggregation, coagulation or flocculation, 
opaque or colored zones called Liesegang structures 
are formed as the diffusive front of $A$ moves ahead
(see Fig.1 for an example).

The precipitation zones that appear in the 
tube usually exhibit a few well defined features. In particular, 
the time of the formation of the $n$-th band, $t_n$, and the 
distance of the $n$-th band from the initial interface, $x_n$,
are related by  the so called {\em time law} \cite{morse}:
\begin{equation}
x_n \sim \sqrt{t_n}
\label{timelaw}
\end{equation}
This law follows evidently from the diffusive nature of 
the reaction-diffusion front.

A more intricate property of the positions of the bands 
is that they usually form a geometric series,
$x_n\sim (1+p)^n$, implying that $x_{n+1}/x_n$, 
tends to a constant for large $n$. This experimental observation is usually 
referred to as the {\it 
Jablczynski law} or {\it spacing law}~\cite{jabli}: 
\begin{equation}
\frac{x_{n+1}}{x_n} \equiv 1+p_n \stackrel{{\scriptstyle n \gg 
1}}{\longrightarrow} 1+p
\label{spacinglaw} 
\end{equation}
The quantity $1+p$ will hereafter be referred to as the {\it spacing 
coefficient}. 
Usually $p>0$ and one speaks of regular banding.

A central goal of the studies of Liesegang bands is 
to understand how the spacing coefficient depends 
on both the experimentally controllable parameters 
(such as the initial concentrations of $A$ and $B$) and 
the other less controllable material parameters such as the diffusion 
coefficients. A major achievement from experimental 
point of view has been the establishment of the so called 
Matalon-Packter law \cite{{Matalon},{Packter}}. It gives 
the dependence of the spacing coefficient on the initial 
concentration, $a_0$, of the outer electrolyte, $A$, in a simple form
\be
1+p=Q_1+\frac{Q_2}{a_0} \ ,
\label{Matpack}
\ee
where $Q_1$ and $Q_2$ depend on the initial 
concentration, $b_0$, of the inner electrolyte as well as 
on all the other material parameters.

As far as the theoretical approaches are concerned, 
several competing theories have been developed
\cite{{Ostwald},{Wagner},{Prager},{zeldo},{Chatter},{shino},{dee},{luthi},{flicker},{kaimul},{ortoleva},{venzl}}, 
many of which 
\cite{{Ostwald},{Wagner},{Prager},{zeldo},{Chatter},{shino},{dee},{luthi}}
fare well in deriving the time and spacing laws \cite{postnuc}. 
They succeed in explaining the formation of these 
bands by following how the 
diffusive reagents $A$ and $B$ turn into immobile precipitate $D$
\be
A+B\rightarrow ... C ... \rightarrow D \quad ,
\label{process1}
\ee
taking into account various scenarios for the intermediate steps 
denoted here by $...C...\,$. The theories are not entirely 
equivalent when they are applied to derive the Matalon-Packter 
law and comparison with experiments slightly favors the
so called {\em induced sol-coagulation} 
model \cite{us}. It should be emphasized, however, that 
there is no clear experimental evidence at present 
that would distinguish decisively among the existing theories. 
Accordingly, it remains to be a task  
to search for distinguishing features of Liesegang phenomena
both in experiments and theories.  

From a chemical point of view the process given 
by (\ref{process1}) often means 
oversimplification. In some cases, for example, the precipitate 
$D$ may partly redissolve in the surplus of component $A$. 
This is not a negligible effect since the precipitation and 
dissolution processes, coupled to 
diffusion, may lead to a propagating 
band  instead of a final static pattern~\cite{miklos_uno}. 

Another effect, that is our main concern in this paper, 
is the dissociation of the invading electrolyte $A$.
In the theoretical approaches formulated up to now, it is 
usually assumed that the concentration of the reacting ions is equal
to the concentration of the electrolyte.  This is  true only
when an ionic compound, a strong acid or a strong base is 
used as $A$. If, however, a weak acid or weak base is used as outer 
electrolyte, then the two concentrations are not equal and the effect of dissociation must be taken  into account. 

As an explicit example, 
imagine the formation of $Mg(OH)_2$ precipitate (this process will
be our main concern throughout this paper). 
Consider a gel soaked with $MgCl_2$, and let us
pour either $NaOH$ or $NH_{4}OH$ onto its surface. In both 
cases the chemistry is the same: $Mg(OH)_2$ precipitate is formed.
There are, however, significant differences in the patterns 
formed as shown in Fig.1. The strong base $(NaOH)$ results in a 
very dense band structure, whereas the weak base $(NH_4OH)$ produces 
a rather loose pattern. A more precise analysis reveals (see below)
that the latter structure has a significantly larger spacing 
coefficient. Based on this example,
one can conclude that either the co-ions or the different degree of 
dissociation have a strong influence on the final pattern. 
In previous experiments on Liesegang phenomena, 
we found that the effects of co-ions (such as $Na^+$ or $NH_4^+$) is 
almost negligible~\cite{miklos_due}. We shall therefore 
assume that the decisive role is played by the dissociation
and that it may be possible to isolate the effect of dissociation 
on the formation of Lieasegang bands. Since we aim at studying this 
effect and since we count physicists in our audience, 
a few words are in order about dissociation.
\end{multicols}
\begin{figure}[htb]
\centerline{
        \epsfysize=0.6\textwidth
        \epsfbox{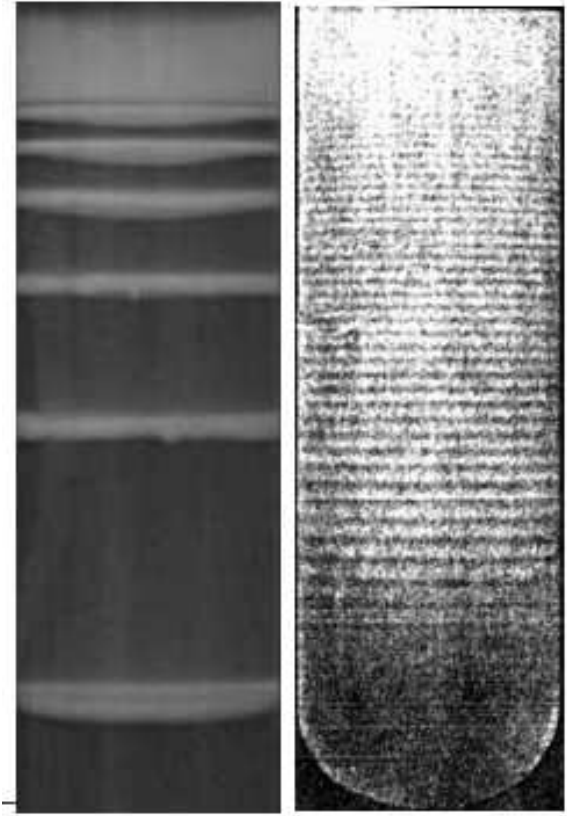}
           }
\vspace{0.2cm}
\caption{$Mg(OH)_2$ Liesegang patterns formed from the reaction of the 
inner electrolyte $MgCl_2$ with two different outer electrolytes:
$NH_4OH$ for the left picture and $NaOH$ for the right one.}
\vspace{0.2cm}
\label{Fig1}
\end{figure}

\begin{multicols}{2}

Dissociation is a common phenomenon 
for acids and bases. Strong acids and bases dissociate 
into ions entirely i.e. the solution 
contains only ions but no neutral molecules. For weak acids 
and bases the dissociation is not complete 
and the neutral 
molecules and ions keep a dynamical equilibrium in the solution.

In order to introduce notation, 
let us consider a weak base ${\cal W}OH$. In an aqueous 
solution, the  base, its cations, ${\cal W}^+$, and 
anions, $OH^-$, are all present
\begin{equation}
{\cal W}OH \quad \rightleftharpoons \quad {\cal W}^+ + OH^- 
\end{equation}
and the equilibrium concentrations are related by
\begin{equation}
K_d=\frac{[{\cal W}^+][OH^{-}]}{[{\cal W}OH]} \label{chimie}
\end{equation}
where $K_d$ is the {\em dissociation constant}.
It is convenient to introduce the degree of dissociation, 
$\alpha$ defined as
\begin{equation}
\alpha=\frac{[{\cal W}^+]}{a_0}=\frac{[OH^-]}{a_0} \label{alpha}
\end{equation}
where $a_0$ denotes the initial concentration of the base, 
$a_0=[{\cal W}OH]_{t=0}$. 
Then eq.(\ref{chimie}) can be written as
\begin{equation}
K_d= a_0 \frac{\alpha^2}{1-\alpha} \label{kd}
\label{Kd}
\end{equation}
and one can see that $\alpha$ can be determined once $a_0$ and
$K_d$ are known.

In order to study the effect of dissociation on Liesegang 
pattern formed by $Mg(OH)_2$ precipitate,  several  bases 
having different values of $K_d$ are required. Unfortunately, 
there are only few inorganic ones available. 
Consequently, we have chosen organic compounds to provide 
the necessary $OH^-$ ions. The organic molecules we shall 
use dissociate like the 1-buthylamin:
\begin{equation}
          C_4 H_{9}-NH_2   +  H_{2}O   \rightleftharpoons
          C_4H_{9}-NH_{3}^+  +  OH^- \ .
\label{chim}
\end{equation}
For our purposes, an important characteristic of an amine 
is its basicity that is conveniently expressed by the 
$pK_a$ of its conjugate acid. For the sake of simplicity, however, 
we shall characterize the dissociation ability of amines by the 
basicity constant $pK_b=14-pK_a$ that can be expressed through $K_d$ as
\begin{equation}
pK_b=-\log{(K_d/e_0)} 
\end{equation}
where $e_0=1{\rm mol/dm^3}$ and $K_d$ is given by eq.~(\ref{Kd}).
The basicity constant is essentially a measure of an amine's
ability to accept a proton from water according to (\ref{chim}). 
The higher the $pK_b$ of an amine the lower is its dissociation ability.

The goal of this work can be now stated more precisely.
We carry out a combined experimental and theoretical 
study of the effect of dissociation on the Liesegang patterns by 
investigating the $pK_b$-dependence of the spacing coefficient 
$1+p$. 

The paper is organized as follows. In Sec.2, we describe the  
experimental procedures and present the experimental data. 
The main finding here is that increasing the dissociation 
constant results in a decreasing spacing coefficient.
In Sec.3, the theoretical models developed 
for explaining the formation of Liesegang patterns are generalized by 
including the dissociation effects. Then  
the spacing coefficient as a function of the dissociation
constant is calculated explicitly and 
comparison with the experimental data follows.
The conclusion is that all
theoretical models that produce the Matalon-Packter law 
should be in qualitative agreement with the experiments.

\section{Experimental part}

\subsection{Materials}

We have studied the formation of  $Mg(OH)_2$ precipitate, by diffusion of  
inorganic and organic compounds  into chemically cross-linked  polyvinylalcohol 
(PVA) hydrogel containing the $MgCl_2$ inner electrolyte. 
\end{multicols}

\begin{center}
\begin{tabular}{|c|c|c|}
\hline
\hspace{1cm} Outer electrolyte \hspace{1cm} &\hspace{1cm} $pK_b$ 
\hspace{1cm} & \hspace{1.5cm} $1+p$ \hspace{1.5cm} \\ \hline \hline
$NH_4OH$ & 4.75 & 1.76 $\pm$ 0.05 \\ \hline
2-buthylamine & 3.44 & 1.13 $\pm$ 0.01 \\ \hline
i-propylamine & 3.40 & 1.110 $\pm$ 0.005\\ \hline
1,2-ethyldiamine & 3.29 & 1.09 $\pm$ 0.01\\ \hline
1-buthylamine & 3.23 & 1.09 $\pm$ 0.01 \\ \hline
piperidine & 2.88 & 1.043 $\pm$ 0.004 \\ \hline
NaOH& $-\infty$ & 1.00 $\pm$ 0.01 \\ \hline
\end{tabular}
\vspace{0.5cm}
\end{center}

Table I. Outer electrolytes, their basicity constants, and 
the corresponding spacing coefficients obtained from the experiments. 
Note that the case of $NaOH$ appears to be special in that 
$x_{n+1}/x_n\rightarrow 1$ for large $n$ and the asymptotic 
form of  $x_n$ should be different from $x_n\sim (1+p)^n$.

\begin{multicols}{2}
Table I lists the 
bases used for producing the ions $OH^-$ as well as their basicity 
constants~\cite{miklos_three}. 
When the spatially separated reactants come into 
contact, the white precipitate $Mg(OH)_2$ is formed as 
a result of the following reaction:
\begin{equation}Mg^{2+ }+ 2OH^- \to  \underline {Mg(OH)_2} \quad .
\end{equation}

\noindent
\subsection{ PVA gels as reaction-diffusion media}

The highly swollen PVA-hydrogels were prepared by cross-linking 
of primary PVA-chains with glutaric aldehyde (GDA) in aqueous 
solution, containing dissolved $MgCl_2$. Commercial PVA 
(Merck 821038) and solution of 25 mass$\%$ GDA (Merck) 
were used for preparation. The initial polymer concentration, 
as well as the cross-linking density has been kept constant. 
The polymer content of the PVA gels was in every case 3.5 mass$\%$. 
The ratio of monomer unit of PVA (VA) to 
the cross-linking agent, GDA was: (VA)/(GDA)= 250. 
In order to study the precipitation and band formation in swollen 
networks, one of the reactants, the inner electrolyte $MgCl_2$ 
was mixed with the polymer 
solution containing the cross-linking agent. 
The concentration of $MgCl_2$  in the gel phase was constant 
($0.05 {\rm mol/dm^3}$). The gelation was induced by 
decreasing the pH of the system by $HCl$ (Reanal, Hungary). 
Then the solution was poured into glass tubes. The gelling process 
usually took 3-4 hours. Experiments 
were made in tubes of length 300 mm and diameter 12 mm.
The tubes were sealed and allowed to stand undisturbed at 
constant temperature of $17^\circ C$. After completion of the 
network formation, the outer reactants were 
brought into contact with the gels, so that they start to 
diffuse into it. The diffusion took place vertically downward 
at controlled temperature. We 
maintained the constancy of boundary condition by refreshing the 
upper solution continuously, thereby keeping the concentration 
of the outer electrolyte at a fixed value (7 mol/l). 
During all the experiment process, constant temperature of $17.00 
\pm 0.02^\circ C$ was maintained and the tubes were kept free 
from mechanical disturbances such as vibrations.
\vspace{0.5cm}

\subsection{Liesegang band formation }

The penetration of  organic as well as inorganic molecules into 
the gel containing the $Mg^{2+}$ ions resulted in precipitate 
bands (see Fig. 1) with a sharp 
interfaces of $Mg(OH)_2$. 
In order to determine the position of bands we used a digital 
video system. A CCD camera with an 1/3" video chip has been 
connected to a PC through a real-time video digitizer card. 
We have also used a scanner to digitalize the 
experimental results. We have determined the coordinate of the 
gel surface where the diffusion started. Then the position of 
the $n$-th band, $x_n$, was measured as the distance between the 
surface of the gel and the upper side of the $n$-th precipitate.


\end{multicols}
\begin{figure}[htb]
\centerline{
        \epsfysize=0.45\textwidth
        \epsfbox{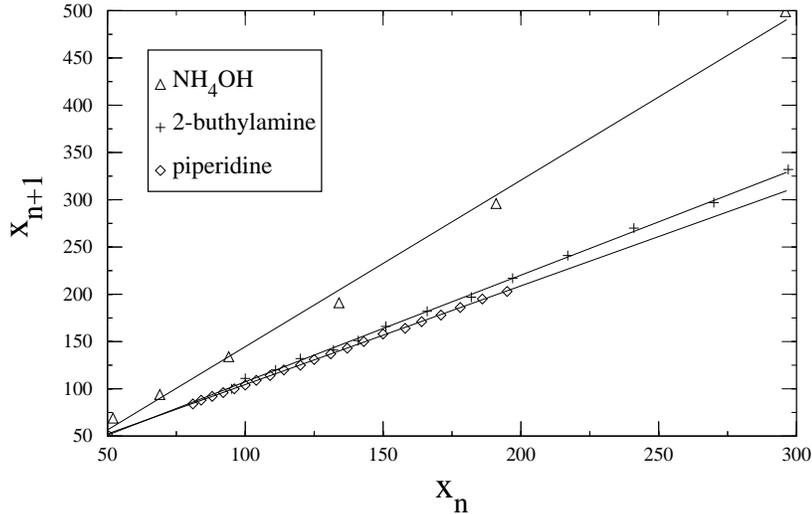}
           }
\vspace{0.5cm}
\caption{Determination of the spacing coefficient given by slope of the curves 
$x_{n+1}$ vs. $x_{n}$ for different compounds. The distances are measured in pixels of the digitalized pictures. The choice of the unit length is irrelevant as we are interested in the ratio $x_{n+1}/x_n$.
The straight lines are least-square fits.}
\label{Fig.2}
\end{figure}
\begin{multicols}{2}
In order to determine the value of the spacing coefficient we plotted
$x_{n+1}$ against $x_n$. For the sake of a  better 
visibility, only three samples are shown on Fig. 
2 but the other sets of data are of similar quality. 
The data were analyzed by linear least square method and 
the resulting spacing coefficients and related errors are 
summarized in Table I.  
Note that the relatively large number of 
bands allowed a rather accurate determination of the 
spacing coefficients (of course, one should 
keep in mind that the errors quoted in Table I are the statistical 
errors and they do not include possible systematic errors).

The $pK_b$ dependence of the spacing coefficient is shown in Fig.3.
One can easily see  
that increasing the dissociation ($pK_b=10^{-K_d}$) decreases
the spacing coefficient. 
It should be pointed out, however, that in contrast to the 
theories discussed below, where all parameters except $K_d$ 
can be kept fixed, the situation is more complex in 
the experiments. All the parameters such as reaction thresholds 
and diffusion constants may, in principle, 
change when going from one outer electrolyte to the other. 
Thus Fig.3 displays the coupling between the spacing coefficient 
and the value of $pK_b$ under the assumption that one can neglect 
the effect of the co-ions on the material parameters that are 
relevant in the process of pattern formation.

\end{multicols}
\begin{figure}[htb]
\centerline{
        \epsfysize=0.5\textwidth
        \epsfbox{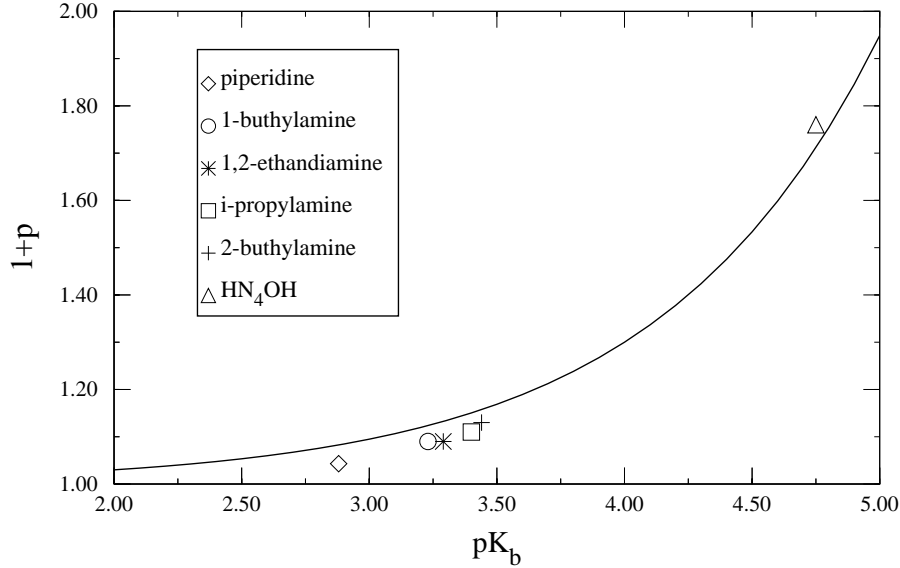}
           }
\vspace{0.5cm}
\caption{Spacing coefficient $1+p$ as a function of the basicity $pK_b$ of the 
invading electrolytes. The full line is the prediction of the theory, 
eq.(22), with parameters $Q_1=1$ and $Q_2=0.003$.}
\label{Fig.3}
\end{figure}

\begin{multicols}{2}

\section{Theoretical approaches}

The formation of precipitation bands is a rather complex phenomenon 
and it is by no means certain that there exist a unified 
description of
all Liesegang phenomena. Indeed, several theories have been developed.
All of them follow how the 
diffusive reagents $A$ and $B$ turn into immobile precipitate $D$
through some intermediate steps 
$A+B\rightarrow ... C ... \rightarrow D$.
The common feature of the theories is
that the precipitate appears as the system goes through some 
supersaturation and nucleation thresholds. The differences 
arise in the details of treating the
thresholds and in the growth kinetics of the precipitate.

Our aim here is not to review these theories in detail 
(a recent critical discussion can be found in~\cite{us}). 
Instead, we shall recall them briefly, highlight their 
specific features and then discuss how 
dissociation enters into the problem. We shall see that
the experimentally relevant regime of dissociation that is fast 
compared to diffusional relaxation, can be treated independently
of the details of the theories. The final result is that,
since the Matalon-Packter follows from
all of the theories, the consequence of the dissociation is 
universal. Namely, the spacing coefficient depends exponentially 
on the basicity constant, $pK_b$.

\subsection{Supersaturation of ion-product theory}

The simplest theory is 
based on the concept of {\it supersaturation of 
ion-product}~\cite{Ostwald}
and has been developed by many researchers
\cite{{Wagner},{Prager},{zeldo},{Smith}}. In this theory, 
the $A,B$ reagents turn into the precipitate $D$ (without any 
intermediate steps) provided the local product of concentrations 
of the reactants, $ab$, reaches some critical value, $q^*$. 
The nucleated particles grow, deplete $A$ and $B$ in their 
surroundings and, as a consequence, nucleation stops. 
When the reaction zone (where $ab$ is maximum) 
moves far enough so that the depletion effect of the 
precipitate becomes weak, then $ab=q^*$ is reached again 
and nucleation can occur.
This quasiperiodic process leads to the formation of successive bands. 
In the limit of very large precipitation- and aggregation rates, the 
control parameters of this model are the initial densities of the 
electrolytes $a_0$ and $b_0$, their diffusion coefficients 
$D_a$ and $D_b$ and the ion-product threshold
$q^*$~\cite{us}.

\subsection{Nucleation-and-growth theory}

The {\em nucleation-and-growth theory} introduces a single intermediate
step in $...C...$ with a mechanism of band formation based on the
supersaturation of the intermediate compound $C$~\cite{dee,luthi}.  In
this theory, $A$ and $B$ react to produce a new diffusing species $C$ 
the nature of which is not really specified.
It may be a molecule as well as a colloid particle. The main event is
the nucleation that occurs when the local
concentration of $C$-s reaches some threshold value. 
The nucleated particles ($D$-s) act as aggregation seeds and the
nearby $C$-s aggregate to the existing droplet (hence become
$D$-s) provided their local concentration is larger than a given
aggregation threshold.  These models are characterized by two
thresholds, $c^*$ for nucleation and $g^*$ for droplet growth. 
As before, $a_0, b_0, D_a, D_b$ are control parameters 
but the diffusion constant $D_c$ of the $C$ species 
and $c^*, g^*$  appear as extra control-parameters~\cite{us}. 
The depletion
mechanism around an existing precipitation band 
is similar to the one described for the ion-product theory
and it leads to the quasiperiodic band formation.


\subsection{Induced sol-coagulation theory}

The nucleation-and-growth theories have been modified to take into account 
effects of the concentration of the electrolytes on the nucleation 
processes. 
In the {\it induced sol-coagulation} theory \cite{{Chatter},{shino}}, 
it is assumed that $A$ and $B$ react to produce a sol ($C$) 
and this sol coagulates if both the concentration of $C$ 
exceeds a supersaturation threshold $c\ge c^*$ and
the local concentration of the outer electrolyte is also 
above a threshold
$a>a^*$. The quantity $a^*$ is often referred 
to as the {\em critical coagulation concentration threshold} 
and is a new free parameter in this theory.
The band formation is a consequence of the nucleation 
and growth of the precipitate combined with the motion of the front
where $a=a^*$.

\subsection{Reaction-diffusion equations}

All the above theories can be described in terms of 
reaction diffusion equations for the concentrations  of the reagents $(a,b,c)$
and of the precipitate $(d)$:
\begin{eqnarray}
\partial_t a =& D_a \partial^2_x a  -&R_1(a,b,c,d)\ ,
\label{prager_uno}\\
\partial_t b =& D_b \partial^2_x b  -&R_1(a,b,c,d)\ , 
\label{prager_due}\\ 
\partial_t c =& D_c \partial^2_x c +&R_2(a,b,c,d)\ ,
\label{prager_tre}\\
\partial_t d =&    &R_3(a,b,c,d)\label{prager_quattro} \ .
\end{eqnarray}
where the reaction terms $R_\alpha$ can always be chosen so that they 
describe the precipitation and aggregation processes which are 
building blocks of the theories discussed above.  
The time- and the spacing laws (\ref{timelaw},\ref{spacinglaw}) 
follow from the equations
corresponding to the above theories. Furthermore, using these 
theories, one can also derive \cite{us}
the experimentally observed Matalon-Packter law (\ref{Matpack}), 
$1+p=Q_1+Q_2/a_0$. This is an important law because $a_0$ is 
one of the few 
experimentally controllable parameter in Liesegang phenomena. 
As we shall see, the knowledge of the simple $a_0$ dependence 
of $1+p$ allows us to 
derive the effects of the dissociation on the spacing coefficient. 

\subsection{Role of dissociation}
 
The outer electrolyte $A$ produces the reagent $\bar A$ through 
reversible dissociation $A\leftrightarrow {\bar A}+A'$. 
This process of dissociation can be characterized
by a relaxation time, $\tau_{dis}$, defined as the
typical time taken by a molecule to dissociate and thus to 
equilibrate the local ionic concentrations. 
This is a {\em microscopic} time and thus it should be 
much smaller than the time of diffusional relaxation of 
density profiles. Indeed, the diffusive front of the $A$ particles 
moves with a velocity $v_{\rm f}\sim \sqrt{D_a/t}$ that diminishes 
with time. Thus the relaxation time of density perturbations over 
a characteristic distance $\ell$ ($\ell$ can be the 
width of the reaction zone,  
$w\sim t^{1/6}$ \cite{GR}, or the distance between consecutive bands
$x_{n+1}-x_n\sim \sqrt{t_n}$) increases without bound i.e. it is a 
{\em macroscopic} time
\begin{equation}
\tau _{diff} \sim \frac{\ell}{v_{\rm f}}\sim t^\sigma \quad , 
\quad \sigma\ge 1/2 \ .
\label{taudiff}
\end{equation}
Consequently, in the long time limit, $\tau_{dis}$ becomes 
negligible and we can assume the existence of a 
{\it local dissociation equilibrium}. Thus we can extend relation 
(\ref{chimie}) to the out-of-equilibrium situation, 
where concentrations depend on $x$ and $t$.  
This implies that, denoting the density of the reagent ion of the 
outer electrolyte $\bar A$ by $\bar a$, 
the first reaction diffusion equation 
(\ref{prager_uno}) must be replaced by the following couple of equations:
\begin{eqnarray}
\partial_t {a} & = & D_{a} \partial_x^2{a} - 
\kappa_1 {a} + \kappa_2 {\bar a}^2 \label{dis_1} \\
\partial_t {\bar a} & = & D_{\bar a} \partial_x^2 {\bar a} + 
\kappa_1 {a} - \kappa_2 {\bar a}^2 - R_1({\bar a},b,c,d) \label{dis_2}\ .
\end{eqnarray}
where $\kappa_1, \kappa_2 \rightarrow \infty$ 
accounting for $\tau_{dis}\gg \tau_{diff}$ and
$\kappa_1/ \kappa_2 = K_d$ ensuring that the steady, homogeneous state
satisfies the steady state condition (\ref{chimie}). The rest of the equations
(\ref{prager_due},\ref{prager_tre},\ref{prager_quattro}) change only by $a \to \bar a$ in the reaction terms.

Equations (\ref{dis_1},\ref{dis_2}) together with the 
rest of the reaction-diffusion equations 
(\ref{prager_due},\ref{prager_tre},\ref{prager_quattro})
can be solved numerically and the spacing coefficient can be determined
within the framework of the various theories. We do not have to 
carry out this work, however, since the concentration of the outer 
electrolyte in a usual experimental 
setup is two orders of magnitude larger than that of the 
inner electrolyte. This means that, for practical purposes, 
the precipitation processes do not influence the concentration profile 
of $A$. In turn, this means that the concentration of ${\bar A}$ 
is determined by 
that of $A$ and, using the fact that the local dissociation equilibrium 
is established fast, we return to the original problem of equations 
(\ref{prager_uno},\ref{prager_due},\ref{prager_tre},\ref{prager_quattro}) 
but with the initial concentration, $a_0$, 
replaced by ${\bar a}_0$ as determined from 
the steady state condition, eq.(\ref{kd}):
\be
{\bar a}_0=\frac{K_d}{2}\left(\sqrt{1+4\frac{a_0}{K_d}}-1\right) \ .
\label{a0bar}
\ee
It follows then that the effect of dissociation on the 
Matalon-Packer law can be obtained by just replacing $a_0$ by ${\bar a}_0$ 
in eq.(\ref{Matpack}):
\be
1+p=Q_1+\frac{2Q_2}
{\sqrt{K_d^2+4a_0K_d}-K_d} \nonumber
\ee
\be
\hskip 46pt =Q_1+\frac{2\cdot 10^{pK_b}Q_2/e_0}
{\sqrt{1+4\cdot 10^{pK_b}a_0/e_0}-1} \ .
\label{Matpack2}
\ee
where we have replaced $K_d$ by the experimentally measured 
basicity parameter $pK_b=-\log(K_d/e_0)$ in the second equation. 
This expression of the
spacing coefficient through the basicity parameter is the central result of 
our theoretical discussion. The derivation has no recourse to the 
details of the underlying theories of precipitation and so the result 
is valid for any theory that reproduces the Matalon-Packter law.

In the experiments discussed in previous chapters, the basicity 
parameter is large ($pK_b\sim  3-5$) and since $a_0\approx 7e_0$, 
we have $10^{pK_b}a_0/e_0\gg 1$. As a consequence, 
the Matalon-Packter law simplifies to 
\be
1+p=Q_1+\frac{Q_2}{\sqrt{a_0e_0}}10^{pK_b/2} \ 
\label{Matpack3}
\ee
thus resulting in a simple exponential dependence on the basicity constant, 
$pK_b$. 

\subsection{Comparison between experiments and theory}

The functional forms obtained above 
(eqs.\ref{Matpack2},\ref{Matpack3}) are universal in the sense that
the $pK_b$ dependence is explicit and only 
$Q_1$ and $Q_2$ depend on the details of the theories.
Unfortunately, $Q_1$ and $Q_2$ contain unknown parameters 
such as aggregation
thresholds and so they should be considered as fitting parameters. 
Nevertheless, it remains a question whether 
the experimental data could be fitted with reasonable values of 
$Q_1$ and $Q_2$ (the only obvious restriction on $Q_1$ and $Q_2$ coming 
from theories is that $Q_1\geq 1$ and $Q_2\geq 0$). 

Fig.3 shows the experimental data together with the theoretical 
curve (\ref{Matpack3}) using $Q_1=1$ and $Q_2=0.003$ and the agreement
we see is very good. This is somewhat surprising. 
Indeed, using various outer 
electrolytes changes not only the dissociation constants but 
may alter the diffusion constants of the 
reagents ($Mg^{2+}$, $OH^-$) as well as  
it may change the various precipitation thresholds. The spacing 
coefficient should depend on all those quantities and the surprise
here is that these dependences are not significant or absent entirely. 
The agreement actually points to the correctness of our initial 
assumption that the $Mg^{2+}$ and $OH^-$ ions dominate the process with the 
co-ions playing no role at all except for the co-ion of $OH^-$ 
setting the initial concentration of the $OH^-$ ions.  

As we can see, the above theoretical studies of dissociation 
and the comparison with experiments 
did not help in deciding which was the right theory for Liesegang phenomena. 
On the other hand, these studies lead to a simple picture 
of how the dissociation affects the spacing coefficient and, furthermore,
it has been possible to express the results in simple analytical form that
should be useful when discussing and designing Liesegang patterns.

\section{Summary}

Our experiments show that the spacing coefficient of Liesegang patterns is 
strongly influenced by the degree of dissociation of the outer electrolyte.  
The dependence of $1+p$ on the basicity constant, $pK_b$ 
has been found to be exponential and 
this experimental finding has been explained on the basis of the 
present theories of Liesegang phenomena.

\section*{Acknowledgments}
We thank T. Antal, P. Hantz, and T. Unger for useful discussions.
This work has been partially supported by the 
Swiss National Science Foundation 
in the framework of the Cooperation in Science and Research with CEEC/NIS,
by the Hungarian Academy of Sciences (Grant OTKA T 019451 and T 015754), and 
by the EPSRC, United Kingdom (Grant No. GR/L58088).

\end{multicols}

\begin{references}
\bibitem{postnuc} There are theories 
                  \cite{{flicker},{kaimul},{ortoleva},{venzl}} 
                  which attribute the formation of Liesegang bands to 
                  the postnucleation instability present
                  in the coarsening process. These 
                  theories however fail in explaining the regularity
                  of the band spacing.
\end{references}
\end{document}